\RequirePackage{lineno}

\documentclass[aps,prc,amsfonts,showpacs,showkeys,nofootinbib,preprintnumbers,superscriptaddress,floatfix]{revtex4-1}

\usepackage{setspace}

\usepackage{rotating}

\usepackage{xfrac}
\usepackage[colorlinks=true, urlcolor=navyblue, linkcolor=navyblue, citecolor=navyblue]{hyperref}

\usepackage{graphicx}
\usepackage{dcolumn}
\usepackage{bm}
\usepackage[printwatermark]{xwatermark}
\usepackage{xcolor}
\usepackage{lipsum}
\usepackage{ulem}

\usepackage{graphicx,amsmath,amssymb}
\usepackage{multirow}
\usepackage{epstopdf}
\usepackage{lineno}
\usepackage{times}
\usepackage{color}

\usepackage{longtable}

\definecolor{navyblue}{rgb}{0,0.08,0.45}

\let\oldeqnarray\eqnarray
\let\oldendeqnarray\endeqnarray

\renewcommand{\emph}[1]
             {{\it #1}} 

\renewenvironment{eqnarray}
  {\linenomathNonumbers\oldeqnarray}
  {\oldendeqnarray\endlinenomath}

\begin{document}

\title{Measurement of the proton spin structure at long distances}

\newcommand*{\ANL}{Argonne National Laboratory, Argonne, Illinois 60439, USA}
\newcommand*{\ANLindex}{1}
\newcommand*{\CANISIUS}{Canisius College, Buffalo, NY 14208, USA}
\newcommand*{\CANISIUSindexindex}{2}
\newcommand*{\CMU}{Carnegie Mellon University, Pittsburgh, Pennsylvania 15213, USA}
\newcommand*{\CMUindex}{3}
\newcommand*{\CUA}{Catholic University of America, Washington, D.C. 20064, USA}
\newcommand*{\CUAindex}{4}
\newcommand*{\SACLAY}{IRFU, CEA, Universit\'{e} Paris-Saclay, F-91191 Gif-sur-Yvette, France}
\newcommand*{\SACLAYindex}{5}
\newcommand*{\CNU}{Christopher Newport University, Newport News, Virginia 23606, USA}
\newcommand*{\CNUindex}{6}
\newcommand*{\UCONN}{University of Connecticut, Storrs, Connecticut 06269, USA}
\newcommand*{\UCONNindex}{7}
\newcommand*{\DUKE}{Duke University, Durham, North Carolina 27708-0305}
\newcommand*{\DUKEindex}{8}
\newcommand*{\DUQUESNE}{Duquesne University, 600 Forbes Avenue, Pittsburgh, PA 15282, USA}
\newcommand*{\DUQUESNEindex}{9}
\newcommand*{\Fairfield}{Fairfield University, Fairfield, Connecticut 06824, USA}
\newcommand*{\FERRARAU}{Universit\`a di Ferrara , 44121 Ferrara, Italy}
\newcommand*{\FERRARAUindex}{10}
\newcommand*{\FIU}{Florida International University, Miami, Florida 33199, USA}
\newcommand*{\FIUindex}{11}
\newcommand*{\FSU}{Florida State University, Tallahassee, Florida 32306, USA}
\newcommand*{\FSUindex}{12}
\newcommand*{\GWUI}{The George Washington University, Washington, DC 20052, USA}
\newcommand*{\GWUIindex}{13}
\newcommand*{\ISU}{Idaho State University, Pocatello, Idaho 83209, USA}
\newcommand*{\ISUindex}{14}
\newcommand*{\INFNFE}{INFN, Sezione di Ferrara, 44100 Ferrara, Italy}
\newcommand*{\INFNFEindex}{15}
\newcommand*{\INFNFR}{INFN, Laboratori Nazionali di Frascati, 00044 Frascati, Italy}
\newcommand*{\INFNFRindex}{16}
\newcommand*{\INFNGE}{INFN, Sezione di Genova, 16146 Genova, Italy}
\newcommand*{\INFNGEindex}{17}
\newcommand*{\INFNRO}{INFN, Sezione di Roma Tor Vergata, 00133 Rome, Italy}
\newcommand*{\INFNROindex}{18}
\newcommand*{\INFNTUR}{INFN, Sezione di Torino, 10125 Torino, Italy}
\newcommand*{\INFNTURindex}{19}
\newcommand*{\INFNPAV}{INFN, Sezione di Pavia, 27100 Pavia, Italy}
\newcommand*{\INFNPAVindex}{20}
\newcommand*{\ORSAY}{Universit\'{e} Paris-Saclay, CNRS/IN2P3, IJCLab, 91405 Orsay, France}
\newcommand*{\ORSAYindex}{21}
\newcommand*{\Juelich}{Institute fur Kernphysik (Juelich), Juelich, Germany}
\newcommand*{\Juelichindex}{22}
\newcommand*{\JMU}{James Madison University, Harrisonburg, Virginia 22807, USA}
\newcommand*{\JMUindex}{23}
\newcommand*{\KNU}{Kyungpook National University, Daegu 41566, Republic of Korea}
\newcommand*{\KNUindex}{24}
\newcommand*{\LAMAR}{Lamar University, 4400 MLK Blvd, PO Box 10009, Beaumont, Texas 77710, USA}
\newcommand*{\LAMARindex}{25}
\newcommand*{\MISS}{Mississippi State University, Mississippi State, MS 39762, USA}
\newcommand*{\MISSindex}{26}
\newcommand*{\ITEP}{National Research Centre Kurchatov Institute - ITEP, Moscow, 117259, Russia}
\newcommand*{\ITEPindex}{27}
\newcommand*{\UNH}{University of New Hampshire, Durham, New Hampshire 03824, USA}
\newcommand*{\UNHindex}{28}
\newcommand*{\NSU}{Norfolk State University, Norfolk, Virginia 23504, USA}
\newcommand*{\NSUindex}{29}
\newcommand*{\OHIOU}{Ohio University, Athens, Ohio 45701, USA}
\newcommand*{\OHIOUindex}{30}
\newcommand*{\ODU}{Old Dominion University, Norfolk, Virginia 23529, USA}
\newcommand*{\ODUindex}{31}
\newcommand*{\JLUGiessen}{II Physikalisches Institut der Universitaet Giessen, 35392 Giessen, Germany}
\newcommand*{\JLUGiessenindex}{32}
\newcommand*{\RPI}{Rensselaer Polytechnic Institute, Troy, New York 12180-3590}
\newcommand*{\RPIindex}{33}
\newcommand*{\URICH}{University of Richmond, Richmond, Virginia 23173, USA}
\newcommand*{\URICHindex}{34}
\newcommand*{\ROMAII}{Universit\`{a} di Roma Tor Vergata, 00133 Rome Italy}
\newcommand*{\ROMAIIindex}{35}
\newcommand*{\MSU}{Skobeltsyn Institute of Nuclear Physics, Lomonosov Moscow State University, 119234 Moscow, Russia}
\newcommand*{\MSUindex}{36}
\newcommand*{\SCAROLINA}{University of South Carolina, Columbia, South Carolina 29208, USA}
\newcommand*{\SCAROLINAindex}{37}
\newcommand*{\TEMPLE}{Temple University,  Philadelphia, PA 19122, USA}
\newcommand*{\TEMPLEindex}{38}
\newcommand*{\JLAB}{Thomas Jefferson National Accelerator Facility, Newport News, Virginia 23606, USA}
\newcommand*{\JLABindex}{39}
\newcommand*{\UTFSM}{Universidad T\'{e}cnica Federico Santa Mar\'{i}a, Casilla 110-V Valpara\'{i}so, Chile}
\newcommand*{\UTFSMindex}{40}
\newcommand*{\INSUBRIA}{Universit\`{a} degli Studi dell'Insubria, 22100 Como, Italy}
\newcommand*{\INSUBRIAindex}{41}
\newcommand*{\BRESCIA}{Universit\`{a} degli Studi di Brescia, 25123 Brescia, Italy}
\newcommand*{\BRESCIAindex}{42}
\newcommand*{\GLASGOW}{University of Glasgow, Glasgow G12 8QQ, United Kingdom}
\newcommand*{\GLASGOWindex}{43}
\newcommand*{\YORK}{University of York, York YO10 5DD, United Kingdom}
\newcommand*{\YORKindex}{44}
\newcommand*{\VIRGINIA}{University of Virginia, Charlottesville, Virginia 22904, USA}
\newcommand*{\VIRGINIAindex}{45}
\newcommand*{\WM}{College of William and Mary, Williamsburg, Virginia 23187, USA}
\newcommand*{\WMindex}{46}
\newcommand*{\YEREVAN}{Yerevan Physics Institute, 375036 Yerevan, Armenia}
\newcommand*{\YEREVANindex}{47}
\newcommand*{\SEOULU}{Seoul National University, Seoul 08826, Korea}
\newcommand*{\SEOULUindex}{48}
\newcommand*{\LJUBLJANA}{University of Ljubljana, Slovenia Jo\v{z}ef Stefan Institute, Ljubljana, Slovenia}
\newcommand*{\LJUBLJANAindex}{49}
\newcommand*{\HAMPTONU}{Hampton University, Hampton, Virginia 23669, USA}
\newcommand*{\HAMPTONUindex}{50}

\author{X. Zheng}
\affiliation{\VIRGINIA}

\author{A. Deur}
\thanks{email: deurpam@jlab.org}
\affiliation{\JLAB}
\affiliation{\VIRGINIA}

\author{H. Kang}
\affiliation{\SEOULU}

\author{S.E. Kuhn} 
\affiliation{\ODU}

\author{M. Ripani}
\affiliation{\INFNGE}

\author{J.~Zhang}
\affiliation{\VIRGINIA}

\author{K.P. Adhikari}
\thanks{Now at {\HAMPTONU}}
\affiliation{\ODU}
\affiliation{\JLAB}
\affiliation{\MISS}

\author{S. Adhikari}
\affiliation{\FIU}

\author{M.J.~Amaryan}
\affiliation{\ODU}

\author{H.~Atac}
\affiliation{\TEMPLE}  
  
% ...

\collaboration{The Jefferson Lab CLAS Collaboration}

\begin{abstract} 
 
Measuring the spin structure of protons and neutrons tests our understanding of how they arise from quarks and gluons, the fundamental 
 building blocks of nuclear matter. At long distances the coupling constant of the strong interaction becomes large, requiring non-perturbative 
 methods to calculate quantum chromodynamics processes, such as lattice gauge theory or effective field theories. 
 Here we report proton spin structure measurements from 
 scattering a polarized electron beam off polarized protons. The spin-dependent cross-sections were measured at large distances, corresponding 
 to the region of low momentum transfer squared between 0.012 and 1.0 GeV$^2$. This kinematic range provides unique tests of chiral effective
field theory predictions. Our results show that a complete description of the nucleon spin remains elusive, and call for further theoretical works, e.g. in 
lattice quantum chromodynamics. 
Finally, our data extrapolated to the photon point agree with the Gerasimov-Drell-Hearn sum rule, a fundamental prediction of quantum field theory that relates the anomalous magnetic moment of the proton to its integrated spin-dependent cross-sections.
\end{abstract}

\maketitle

Understanding how hadronic matter arises from its fundamental constituents, quarks and gluons, is central to the study of nuclear and
particle physics. 
Although the strong interaction is described by Quantum Chromodynamics (QCD), it remains the least understood force in the Standard Model. The difficulty arises because 
the QCD coupling constant $\alpha_s$ becomes large at long distances~\cite{Deur:2016tte}, making traditional perturbative expansions in powers of $\alpha_s$ infeasible.
Consequently, complex phenomena like quark confinement 
are hard to understand quantitatively. The most fundamental
approach to
calculate QCD non-perturbatively is lattice gauge theory~\cite{Tanabashi:2018oca}. 
A second approach is provided by Effective Field Theories (EFT), which maintain rigorous, traceable
connections to the underlying fundamental theory. 
A popular approach is chiral effective field theory ($\chi$EFT)~\cite{Bernard:2007zu,Scherer:2009bt}, which is constructed
from hadronic degrees of freedom and incorporates the symmetries of QCD, including its
approximate chiral symmetry. By making use of a perturbative expansion in small parameters, $\chi$EFT 
predicts experimental observables from a limited set of phenomenological inputs. 
Although generally successful, $\chi$EFT has been challenged by experimental data that depend explicitly
on spin degrees of freedom~\cite{Kuhn:2008sy,Deur:2018roz}. 
This is not unprecedented: other
theoretical predictions had been thought to be robust until confronted with spin observables, including
parity symmetry~\cite{Wu:1957my}, the Ellis-Jaffe spin sum
rule~\cite{Ellis:1973kp}, 
the nucleon spin asymmetry $A_1$~\cite{Brodsky:1994kg}, and  
lattice QCD calculations of the nucleon axial charge~\cite{Chang:2018uxx}. 
Therefore, fully understanding QCD and nuclear matter requires an extensive set of spin observables. 

We report on the measurements performed using a polarized electron beam to probe a polarized proton at the  
Thomas Jefferson National Accelerator Facility (Jefferson Lab), in Virginia, USA. 
We measured spin-dependent cross
sections in the nucleon resonance region at very low $Q^2$, i.e. at long distances. 
Here, $Q^2$ is the square of the 4-momentum transferred 
from the electron to the proton and represents the inverse of the distance scale probed by the scattering. 
Polarized electrons with energies of 3.0, 2.3, 2.0, 1.3 and 1.1~GeV, produced by Jefferson Lab's Continuous Electron Beam Accelerator Facility (CEBAF), were scattered from a polarized proton  target~\cite{Crabb:1995xi,Keith:2003ca}. The beam polarization ($P_b$) was measured to be 85\% with a total uncertainty of 2\% using
a M{\o}ller polarimeter~\cite{Mecking:2003zu}. The target contained granules of NH$_3$ that were dynamically polarized~\cite{Crabb:1995xi} at 1K in a 5~T magnetic field.
The target polarization ($P_t$) varied from 75\% to 90\%, as monitored by nuclear magnetic resonance polarimetry. As described below and in the Methods section, the product $P_bP_t$ was measured to a relative precision of $(2-5)\%$. 
The scattered electrons were identified using the CEBAF Large Acceptance Spectrometer (CLAS)~\cite{Mecking:2003zu}, which was equipped with a multi-layer drift chamber detector for charged particle tracking, a scintillator hodoscope for particle time-of-flight measurement, an electromagnetic calorimeter and a 
Cherenkov Counter for discriminating scattered electrons from other background particles. The Cherenkov Counter in one of the six sectors of CLAS was modified specifically for this experiment to detect electron scattering at angles as low as $6^\circ$. Only this sector was used to collect the inclusive electron scattering data reported here.

\begin{figure}[!h]
\includegraphics[width=0.35\textwidth]{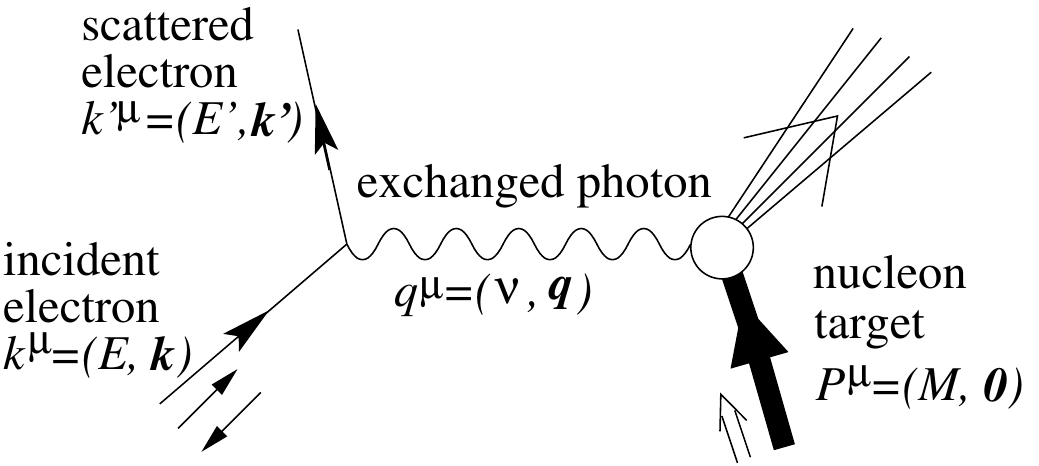}
\caption{The one-photon exchange process   in electron-nucleon scattering. 
The 4-momenta of the   photon, the
  incident and the scattered electrons are 
    $q^\mu =(\nu,\mathbf{q})$,
  $k^\mu =(E,\mathbf{k})$ and $k'^\mu =(E',\mathbf{k'})$, respectively. The spin direction
  of the incident electron is indicated by the arrows $\uparrow\downarrow$.
 The nucleon, if at rest, has $P=(M,\mathbf{0})$ and its spin is indicated by the outlined arrow $\Uparrow$. }\label{fig:escatt}
\end{figure}

The dominant scattering process is the one-photon exchange, in which the incident electron exchanges a single virtual photon with the nucleon of mass $M$, see Fig.~\ref{fig:escatt}. 
The 4-momentum transferred from the electron to the nucleon is $q^\mu=k^\mu-k'^\mu=(\nu,\mathbf{q}$), in which $k^\mu$ and $k'^\mu$ are the 4-momenta of the incident and the scattered electrons, respectively, and $\nu$ is the energy transfer. 
In the following, we describe this process using 
the Lorentz-invariant variables
$Q^2=-q^2$, and the Bjorken scaling variable 
$x\equiv - q^2/(2P\cdot q)$ or the invariant
mass of the photon-nucleon system 
$W\equiv \sqrt{(P+q)^2} = \sqrt{P^2 + (1/x-1)Q^2}$. 
The inclusive electron scattering cross section can be written as a linear combination of structure
functions, of which $F_1(x,Q^2)$ and $F_2(x,Q^2)$
represent the spin-independent part of the cross section, and 
  the spin structure functions
$g_1(x,Q^2)$ and $g_2(x,Q^2)$ 
describe its dependence on the beam and target spin polarization.
These structure functions encode the internal structure of the target. 
Alternatively, one can describe the spin-dependent part of the nucleon response in terms of virtual photo-absorption asymmetries $A_1=[g_1-(Q^2/\nu^2) g_2]/F_1$ and $A_2=(\sqrt{Q^2}/\nu)(g_1+g_2)/F_1$~\cite{Fersch:2017qrq}. 
The polarized cross section difference  
$\Delta \sigma \equiv \sigma^{\downarrow\Uparrow} - \sigma^{\uparrow\Uparrow}$, with $\uparrow\downarrow$ representing the beam helicity state and $\Uparrow\Downarrow$ the target spin orientation, 
is largely proportional to $g_1$ (or equivalently $A_1 F_1$) with a small contribution from $A_2 F_1$. 
 
The proton spin structure function $g_1$ and the product $A_1F_1$ were 
extracted from the difference in the measured yield, $N$, of scattered electrons from a longitudinally polarized target between opposite beam helicity states: 
\begin{eqnarray}
\frac{N^{\downarrow\Uparrow}}{Q_b^{\downarrow}} - \frac{N^{\uparrow\Uparrow}
}{Q_b^{\uparrow}}
= \Delta \sigma(W, Q^2){\mathcal{L} P_b P_t  a(W, Q^2)},\label{eq:polyield}
\end{eqnarray}
where $Q_b$ is the time-integrated beam current, 
$\mathcal{L}$ is the areal density of polarized protons
in the target, and $a(W, Q^2)$ accounts for the detector acceptance and efficiency.
The product $\mathcal{L} P_b P_t$ was measured directly using elastic
scattering on the proton 
and $a(W, Q^2)$ was determined using a Monte Carlo simulation of
the experiment; see the Methods section for details. 
Examples of our $g_1$ results on the proton are shown in Fig.~\ref{Fig:g1}.
The full data set is given in the Supplemental Material.
\begin{figure}[htp]
\includegraphics[width=0.5\textwidth]{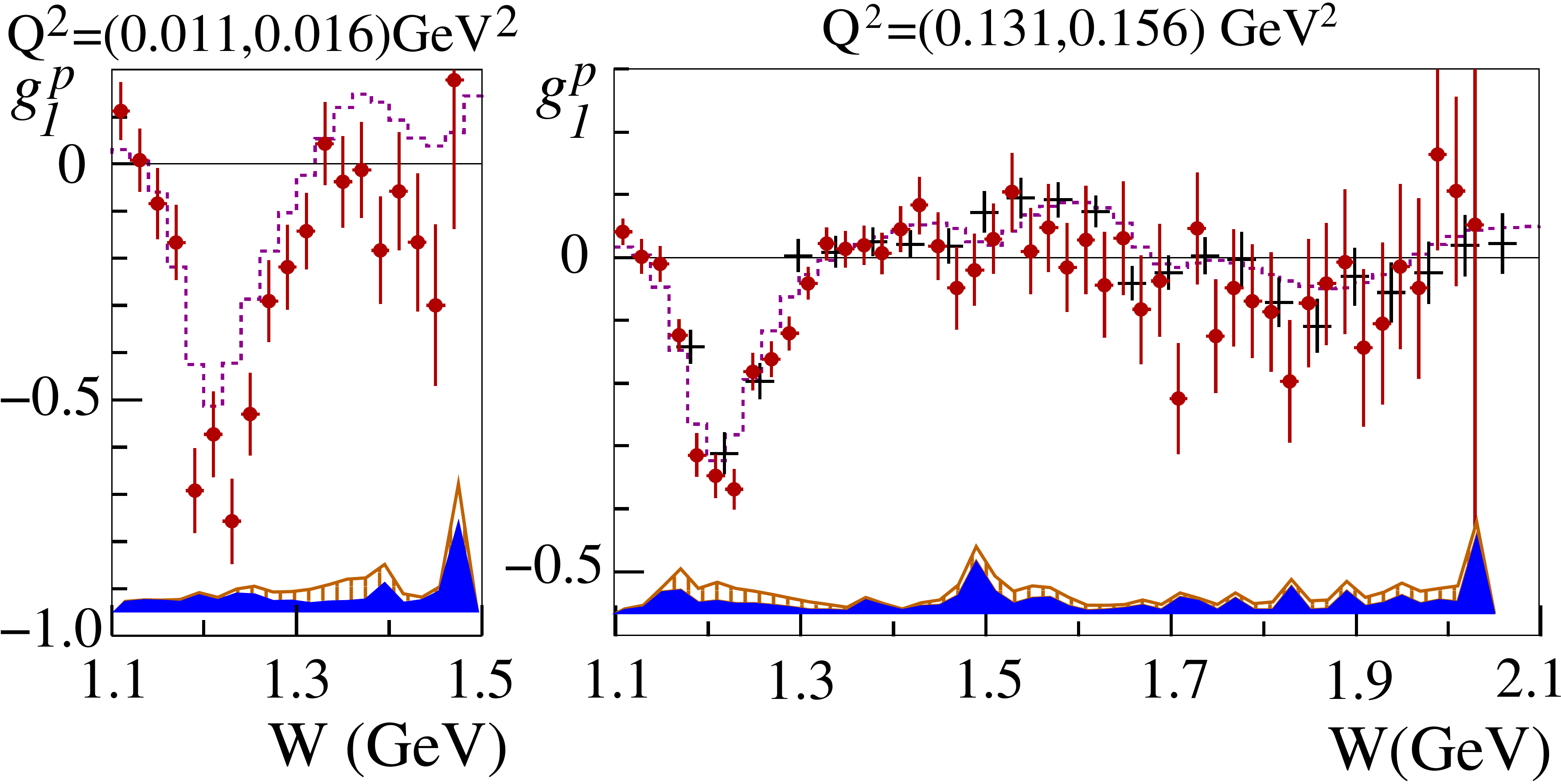}
\caption{
Results on $g_1(Q^2,W)$ of the proton.   Data from this work (solid circles) are plotted vs invariant mass $W$ for the lowest
  ($0.011 \leq Q^2 \leq 0.016$~GeV$^2$) bin and
  an intermediate ($0.131\leq Q^2\leq 0.156$~GeV$^2$) bin, compared to
  a parameterization of previous world data (dotted curve)~\cite{Fersch:2017qrq}.
  The error bars are statistical. The solid and the vertically-hatched bands show the 
  experimental and the parameterization uncertainties, respectively.   The horizontal line is the zero
  of the vertical axis. Results from
  a previous experiment carried out in Jefferson Lab's Hall B~\cite{Fersch:2017qrq} are
  shown when available (crosses), with the error bars representing the statistical and systematic 
 uncertainties added in quadrature. 
}
\label{Fig:g1}
\end{figure}
Our results extend the measured $Q^2$ range down to 
  below the pion mass squared ($m_\pi^2$), three times smaller than previous data~\cite{Amarian:2002ar,Amarian:2003jy,Amarian:2004yf,Deur:2004ti,Deur:2008ej,Dharmawardane:2006zd,Prok:2008ev,Guler:2015hsw, Fersch:2017qrq}, 
which makes 
it possible to rigorously test $\chi$EFT calculations for spin-dependent observables.

In our study, we utilize sum rules that relate integrals of structure functions 
to amplitudes calculable by lattice QCD~\cite{Chambers:2017dov,Liang:2019frk} or $\chi$EFT, or to known static properties of the target.
One such relation is the Gerasimov-Drell-Hearn 
(GDH) sum rule~\cite{Gerasimov:1965et,Drell:1966jv}
for real photon absorption ($Q^2=0$):
\begin{eqnarray}
\int_{\nu_{0}}^{\infty} \Delta \sigma(\nu) \frac{d\nu}{\nu} &=&  - \frac{2\pi^2\alpha}{M^2}\kappa^2,
\label{eq:gdh}
\end{eqnarray}
with $\kappa$ the anomalous magnetic moment of the target particle, $\nu_{0}$ the inelastic threshold and $\alpha$ the fine-structure constant.
Theoretical arguments 
indicate that the divergence of the $1/\nu$ factor is compensated by the fast decrease of $\Delta\sigma$ with $\nu$. This is supported by experiments which
have verified the GDH sum rule for the proton within about 7\% accuracy~\cite{Dutz:2004zz,Hoblit:2008iy}. 
There exist several prescriptions that generalize 
the GDH sum rule to electron scattering
  in terms of moments of spin structure functions 
integrated over $x$ (which is equal to $Q^2/2M\nu$ in the laboratory frame).
One often-used generalization is~\cite{Ji:1999mr}:
\begin{eqnarray}
\Gamma_1(Q^2) &\equiv& \int_0^{x_0}g_1(x,Q^2)dx= \frac{Q^2}{2M^2} I_1(Q^2),
\label{eq:gdhsum_def2}
\end{eqnarray}
where $x_0 =Q^2/(W_{thr}^2-M^2+Q^2)$ corresponds to the electroproduction threshold
$W_{thr} = M + m_{\pi}$   = 1.073 GeV. 
Equation~(\ref{eq:gdhsum_def2}) defines the integral $I_1$, which is related to the first polarized doubly-virtual Compton scattering (VVCS) amplitude that is calculable in the $\nu \to 0$ limit with lattice QCD or $\chi$EFT~\cite{Bernard:2007zu,Scherer:2009bt, Bernard:1992nz, Ji:1999pd,Ji:1999sv, Bernard:2002bs,Bernard:2002pw, Kao:2002cp, Bernard:2012hb,  Lensky:2014dda,Lensky:2016nui, Alarcon:2020icz}.
The other prevailing generalization of the GDH integral is~\cite{Drechsel:2002ar}: 
\begin{eqnarray}
    I (Q^2)   &=&  \frac{2M^2}{Q^2} \int_0^{x_0}\left[A_1(x,Q^2) F_1(x,Q^2)\right]dx,
\label{eq:gdhsum_def1}
\end{eqnarray}
which can be calculated from both the first and the second spin-dependent VVCS amplitudes in the $\nu\to 0$ limit. 
The $ I (Q^2)$ thus obtained can be extrapolated to $Q^2 = 0$ to test the
original GDH prediction $ I (0) = {\kappa^2}/{4}$.  
In this work, we present results on both generalizations.

To form the spin structure integrals in Eqs.~(\ref{eq:gdhsum_def2} \& \ref{eq:gdhsum_def1}), the
measured values of $g_1$ or $A_1F_1$ were used whenever available from our experiment up to a maximum $x$ corresponding to $W=1.15$~GeV, which was chosen to limit the background from the elastic radiative tail (see Methods section)  and down to a minimum $x$ 
determined by the beam energy and the acceptance of CLAS.
Contributions from regions at low $x$ (down to $x=10^{-3}$) and at high $x$
from $W_{thr}$ to $W=1.15$~GeV 
were evaluated using a parameterization of previous data~\cite{Fersch:2017qrq}. 

\begin{figure}[htp]
  \includegraphics[width=0.48\textwidth]{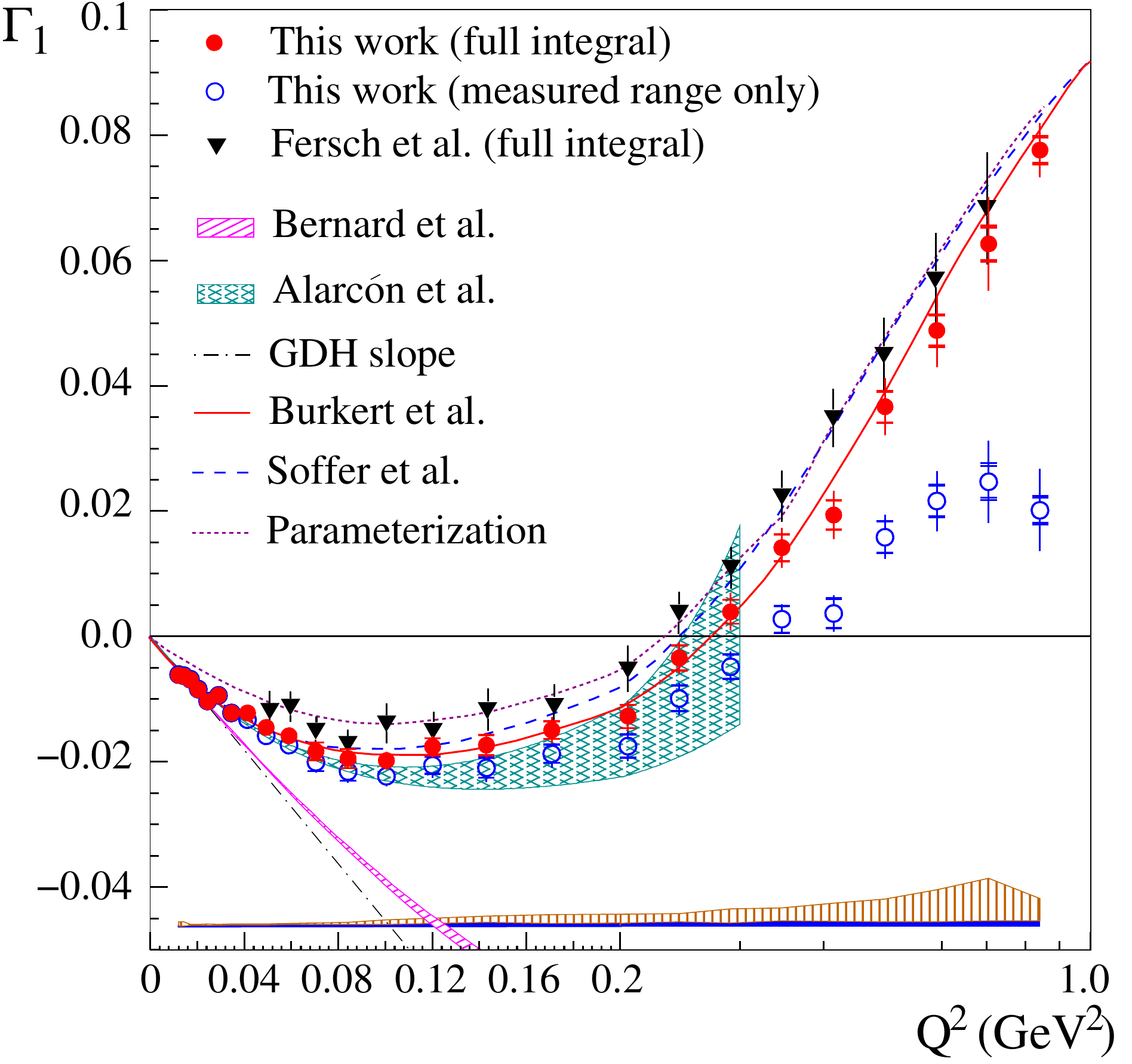}
\caption{Results on $\Gamma_1(Q^2)$ for the proton. Integrals over the experimentally covered $x$ range are shown
  as open circles. Full integrals are shown as solid circles. 
  The inner and the outer error bars (sometimes too small to be seen) are for statistical and total uncertainties, respectively. Results
  from a previous experiment~\cite{Fersch:2017qrq} are shown as solid triangles   whose error bars display the statistical and systematic 
 uncertainties added in quadrature. The  solid and the vertically-hatched bands show 
  the experimental and the parameterization uncertainties, respectively. 
  Also shown are the latest $\chi$EFT predictions
  by Bernard {\it et al.}~\cite{Bernard:2012hb} (diagonally hatched band) and Alarc\'on {\it et al.}~\cite{Alarcon:2020icz} (cross-hatched band),
  phenomenological models by Burkert {\it et al.}~\cite{Burkert:1993ya} (solid curve)
  and Soffer {\it et al.}~\cite{Pasechnik:2010fg} ( dashed curve), as well as our
  spin structure function parameterization~\cite{Fersch:2017qrq} (dotted curve). The dash-dotted line is the slope predicted by the GDH sum rule as $Q^2 \to 0$.}
\label{Fig:GDH_P}
\end{figure}
\begin{figure}[!ht]
\includegraphics[width=0.45\textwidth]{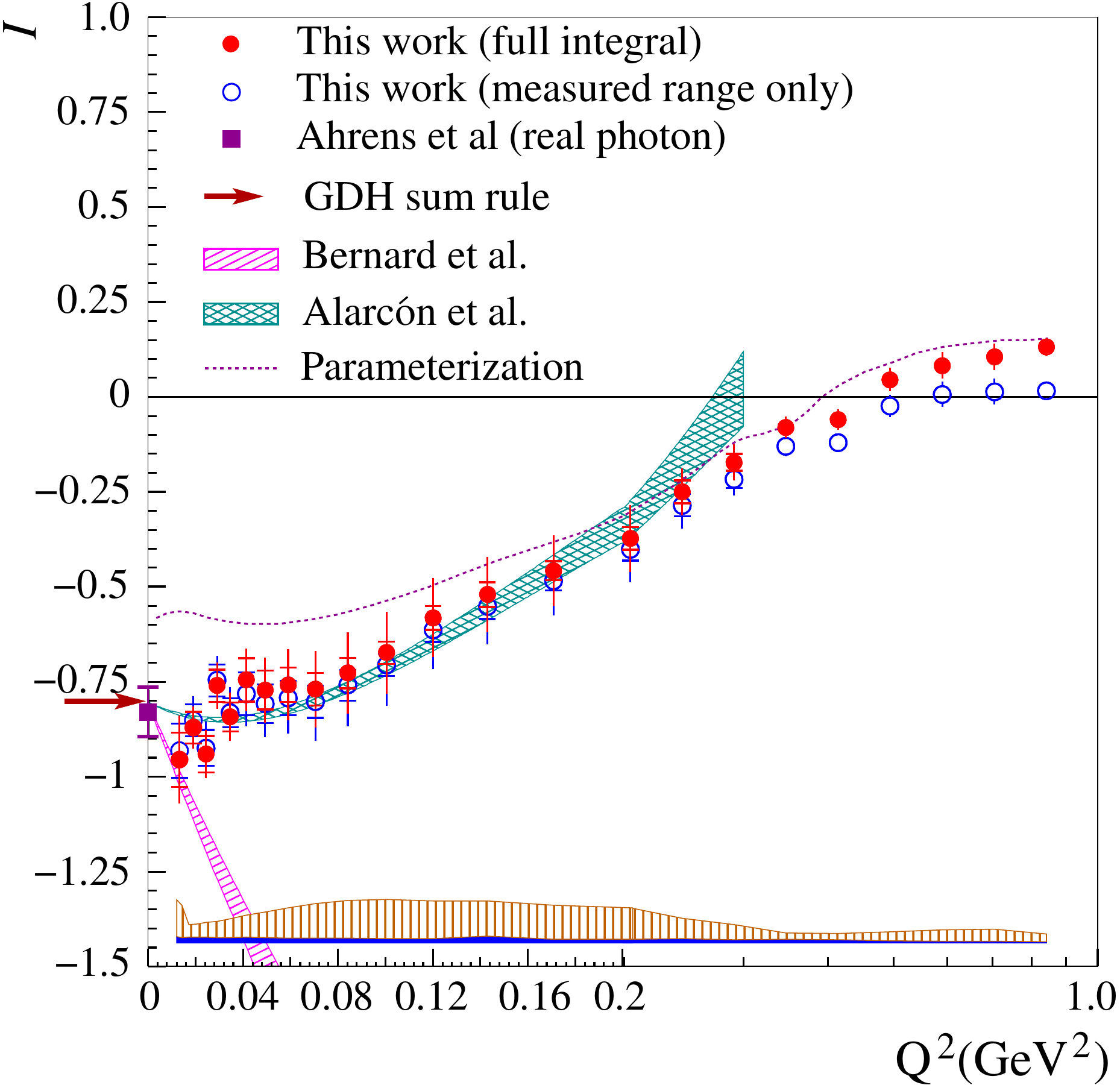}
\caption{Results on $I(Q^2) $ for the proton.  
Integrals over the experimentally covered $x$ range are shown
  as open circles. Full integrals are shown as solid circles. 
  The inner and the outer error bars (sometimes too small to be seen) are for statistical and total uncertainties, respectively.  The  solid and the vertically-hatched bands show 
  the experimental and the parameterization uncertainties, respectively. 
  Also shown are the latest $\chi$EFT predictions
  by Bernard {\it et al.}~\cite{Bernard:2012hb} ( diagonally hatched band) and  Alarc\'on {\it et al.}~\cite{Alarcon:2020icz} (cross-hatched band),
  and our spin structure function parameterization~\cite{Fersch:2017qrq} (dotted curve). The dash-dotted line is the slope 
  predicted by the GDH sum rule as $Q^2 \to 0$.
The GDH value is shown by the arrow at $I^{GDH}=-0.804$. The experimental photoproduction result~\cite{Dutz:2004zz,Hoblit:2008iy} is shown by the solid square   with the error bar providing the statistical and systematic uncertainties added in quadrature.}
\label{Fig:ITT_P}
\end{figure}

Results on $\Gamma_1(Q^2)$ and $ I (Q^2)$ are shown in Figs.~\ref{Fig:GDH_P} and~\ref{Fig:ITT_P}.
To quantify the degree of agreement between our data and the recent $\chi$EFT predictions~\cite{Bernard:2012hb, Alarcon:2020icz}, we computed the $\chi^2$ per degree of freedom between these predictions and our results. We find that the predictions in~\cite{Bernard:2012hb} agree with our results only at the lowest few $Q^2$ points, up to $Q^2=0.024 (0.014)$~GeV$^2$ for $\Gamma_1$ ($ I $), if we require a $\chi^2_\mathrm{reduced}<2$. On the other hand, the predictions in~\cite{Alarcon:2020icz} agree with our data over their full range, with $\chi^2_\mathrm{reduced}<2$ up to $Q^2 = 0.3$~GeV$^2$. 
The phenomenological models~\cite{Burkert:1993ya,Pasechnik:2010fg} 
agree well with our results for all $Q^2$ values. 
The new results on $\Gamma_1(Q^2)$ generally agree with a previous
experiment~\cite{Fersch:2017qrq} 
in the overlapping $Q^2$ region. However, 
there exist visible differences between our results and 
  the spin structure function
parameterization~\cite{Fersch:2017qrq}, indicating that it can be improved with our data.
Extrapolating our results on $ I (Q^2)$ to $Q^2=0$ yields 
\begin{eqnarray}
   I ^{exp}(0)=-0.798\pm  0.042
\end{eqnarray}
 assuming the $Q^2$-dependence
of $I$ predicted by Alarc\'on {\it et al.}~\cite{Alarcon:2020icz} within their quoted theoretical uncertainty (see details in the Methods section).
This result is in good agreement with the GDH sum rule prediction
$ I ^{GDH}=-{\kappa^2}/{4}=-0.804(0)$ for
the proton and with the experimental photoproduction result 
$-0.832\pm 0.023\mathrm{(stat)}\pm 0.063\mathrm{(syst)}$~\cite{Dutz:2004zz,Hoblit:2008iy}. 
Our results provide a test of the GDH sum independent from exclusive photoproduction~\cite{Dutz:2004zz,Hoblit:2008iy}. 

Predictions from $\chi$EFT for $ I (Q^2)$ and $\Gamma_1(Q^2)$ are constrained at $Q^2=0$ by the GDH sum rule. No such constraint is available for $\gamma_0(Q^2)$, the generalized longitudinal spin polarizability, related by a sum rule to the integral of  $A_1F_1$~\cite{Guichon:1995pu, Drechsel:2002ar}:
\begin{eqnarray}
   \gamma_0(Q^2) =  \frac{16\alpha M^2}{Q^{6}}\int_0^{x_0}x^2 A_1(x,Q^2) F_1(x,Q^2) dx. \label{eq:gamma_0}
\end{eqnarray}
This endows $\gamma_0(Q^2)$ with additional resolving power to test the several theoretical predictions available. Furthermore, the $x^2$ weighting in Eq.~(\ref{eq:gamma_0}) suppresses the low-$x$ contribution. This is beneficial
since the low-$x$ region is inaccessible experimentally and must be estimated using models, which introduces model uncertainty. 
The two integrals $ I $ and $\gamma_0$ have different systematic uncertainties and therefore provide complementary tests of theoretical predictions.

\begin{figure}[!ht]
\includegraphics[width=0.48\textwidth]{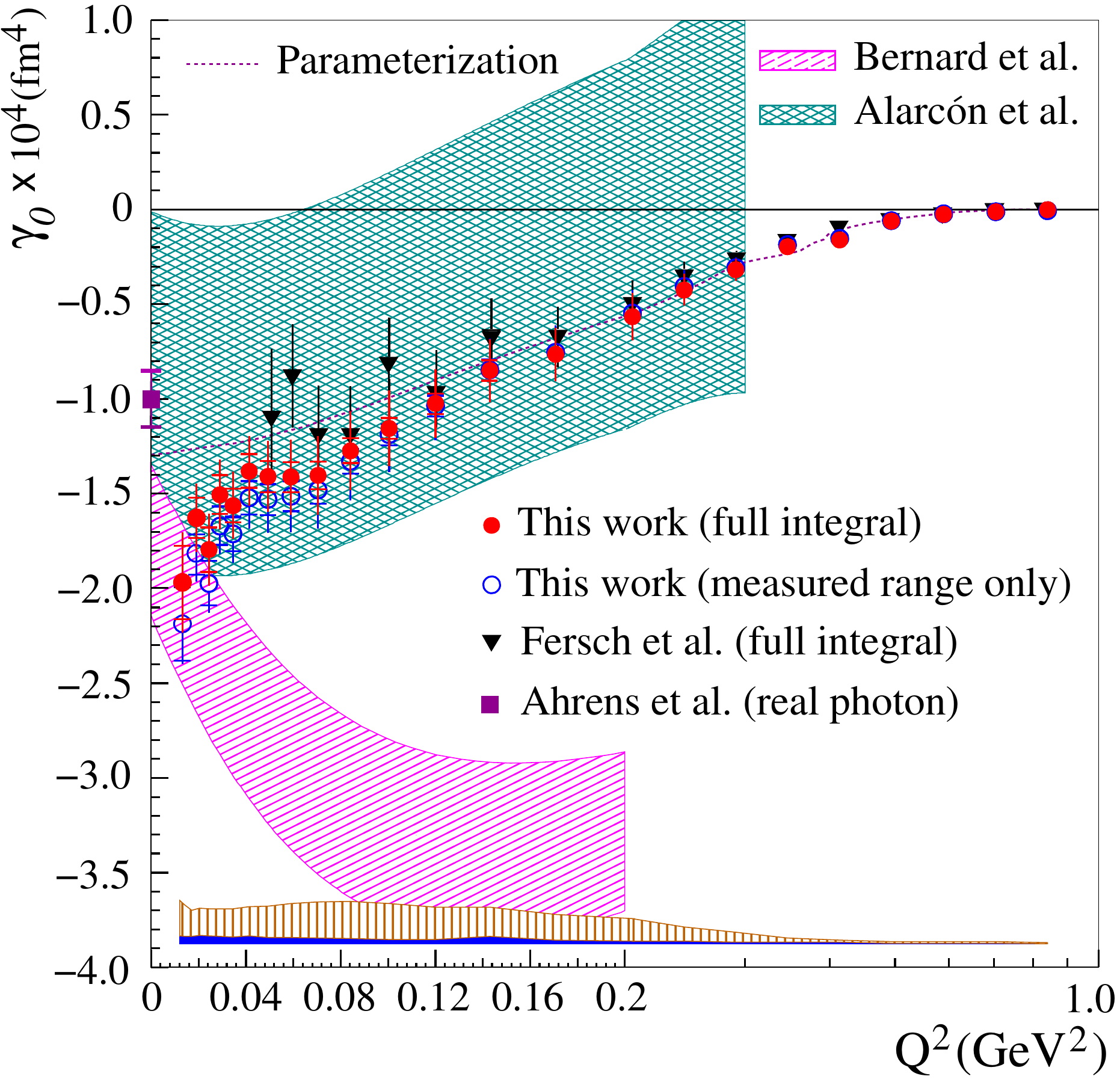}
\caption{Results on ${\gamma_0}(Q^2)$ for the proton.  
Integrals over the experimentally covered x range are shown as open circles. Full integrals are shown
as solid circles. The inner and the outer error bars (sometimes too small to be seen) are for statistical and total uncertainties, respectively. The solid and the vertically-hatched bands show the experimental and the parameterization uncertainties, respectively. Results from a previous experiment~\cite{Fersch:2017qrq} are shown as solid triangles whose error bars display the statistical and systematic uncertainties added in quadrature.
Also shown are the latest $\chi$EFT predictions
by Bernard {\it et al.}~\cite{Bernard:2012hb} ( diagonally hatched band) and  Alarc\'on {\it et al.}~\cite{Alarcon:2020icz} (cross-hatched band),
and our spin structure function parameterization~\cite{Fersch:2017qrq} (dotted curve).
The photoproduction data point~\cite{Dutz:2004zz,Hoblit:2008iy,Gurevich:2017cpt}
is shown as the solid square   with the error bar providing the statistical and systematic uncertainties added in quadrature.}
\label{Fig:gamma0_P}
\end{figure}

Our results for $\gamma_0(Q^2)$ are shown in Fig.~\ref{Fig:gamma0_P}. 
Neither of the new $\chi$EFT calculations describes the full data set well: 
The calculation from Ref.~\cite{Bernard:2012hb} agrees in magnitude (but not in slope) with our lowest $Q^2$ results up to $Q^2 \approx 0.025$~GeV$^2$, while the calculation from Ref.~\cite{Alarcon:2020icz} describes
the shape of   the data only marginally
below that $Q^2$ value.
Together with the photoproduction data point~\cite{Dutz:2004zz,Hoblit:2008iy,Gurevich:2017cpt}, 
our data indicate a strong change in $Q^2$ slope towards a value consistent with that predicted in Ref.~\cite{Bernard:2012hb} at very low $Q^2$.
Classically, $\gamma_0$ represents the distortion of the proton spin structure in response to the interference 
between various transverse electric and magnetic field components of the virtual photon shown in Fig.~\ref{fig:escatt}.  
In a hadronic picture, $\gamma_0$ is principally due to the difference between the contribution from the pion cloud of the proton 
(positive) and the excitation of the $\Delta$ (negative)~\cite{Alarcon:2020icz}. 
The data thus indicate that the $\Delta$ contribution dominates at the photon point and becomes even 
more important for small-$Q^2$ virtual photons. 
This may be pictured intuitively from the extended size of the pion cloud whose contribution is quickly suppressed with increasing $Q^2$. 
However, at higher
$Q^2$, the slope turns over  
since the polarizability is a global feature of the proton which must vanish as $Q^2 \to \infty$, as seen from the $1/Q^6$ factor in Eq.~(\ref{eq:gamma_0}).

Although the upper bound of the validity domain
of $\chi$EFT is not known, the kinematic coverage of our data is well within its expected range between $m^2_\pi \approx 0.02$~GeV$^2$ 
and  the chiral symmetry breaking scale, 
$\Lambda_\mathrm{\chi}^2
\approx 1$~GeV$^2$.
The actual validity range depends on the orders of the expansion parameter $m_\pi / \Lambda_\mathrm{\chi}$ at which the calculations are done, the expansion method, and the observable.
One reason for the limited success of $\chi$EFT in describing our results 
may be coming from the difficulty to fully account for the $\Delta$ resonance, the proton's first excited state.
In fact, early $\chi$EFT calculations~\cite{Bernard:1992nz, Ji:1999pd,Ji:1999sv} did not explicitly include 
the $\Delta$ excitation, which slows down the convergence of the $\chi$EFT perturbation series,
or they included it phenomenologically~\cite{Bernard:2002bs,Bernard:2002pw, Kao:2002cp}.
This was thought to be the reason why many 
of the early nucleon spin structure function data~\cite{Amarian:2002ar,Amarian:2003jy,Amarian:2004yf,Deur:2004ti,Deur:2008ej,Dharmawardane:2006zd,Prok:2008ev,Guler:2015hsw} disagreed with 
calculations~\cite{Bernard:1992nz, Ji:1999pd,Ji:1999sv, Bernard:2002bs,Bernard:2002pw, Kao:2002cp}. 
This disagreement prompted refined $\chi$EFT calculations~\cite{Bernard:2012hb, Lensky:2014dda,Lensky:2016nui, Alarcon:2020icz} and an  
experimental program at Jefferson Lab optimized to cover the $\chi$EFT
domain~\cite{Adhikari:2017wox, Sulkosky:2019zmn}, including the measurement reported here. 
The latest calculations~\cite{Bernard:2012hb,Lensky:2014dda,Lensky:2016nui,Alarcon:2020icz} 
both include the $\Delta$ but differ in their expansion method to account for   the $\pi$-$\Delta$ corrections.
Ref.~\cite{Bernard:2012hb} treats the nucleon-$\Delta$ mass difference $\delta M$ as a small parameter of the same order as $m_\pi$.  Refs.~\cite{Lensky:2014dda,Lensky:2016nui,Alarcon:2020icz} use $\delta M$ as an intermediate scale such that $\delta M/\Lambda_\chi \approx m_\pi/\delta M$ is used as the expansion parameter to account for the $\Delta$. 
  In addition, the calculations~\cite{Lensky:2014dda,Lensky:2016nui,Alarcon:2020icz} include empirical form factors in the relevant couplings to approximate the expected impact of high-order contributions. They make $\gamma_0$ vanish at large $Q^2$,
as observed, in contrast to calculation~\cite{Bernard:2012hb}
which purely contains terms computed with $\chi$EFT
and has no free-parameter that can be adjusted.
For $\gamma_0$, which arises at third order in the $\pi$-N loops 
there are large cancellations between $\pi$-N loops and the $\Delta$ contribution. 
Therefore, the calculations are very sensitive to the expansion and renormalization scheme, and the order of the expansion. This is 
why $\gamma_0$ is especially well-suited to test $\chi$EFT.
 Finally, the integrals $\Gamma_1$ and $I$ contain Born terms in addition to the polarizability contributions calculated in $\chi$EFT. These terms are constrained by the GDH sum rule at $Q^2 = 0$. Refs~\cite{Lensky:2014dda,Lensky:2016nui,Alarcon:2020icz} assume that their $Q^2$-dependence  follows the corresponding proton form factors. This $Q^2$-dependence leads to the difference with Ref.~\cite{Bernard:2012hb} and the agreement with our data.
 We remark that the shaded theory bands in Figs.~\ref{Fig:GDH_P}-\ref{Fig:gamma0_P} parameterize some of the uncertainties specific to each theoretical calculation, which are different for the two approaches. We refer the reader to the original publications~\cite{Bernard:2012hb, Alarcon:2020icz} for details.

Although it is essential to understand the fundamental forces of nature from first
principles, such descriptions are often  impossible and one must use effective
theories based on the new degrees of freedom that emerge from 
complexity~\cite{Anderson:1972pca}.
   The leading effective theory for the strong interaction, $\chi$EFT, has been precisely tested by
our very low $Q^2$ measurement of $\Gamma_1$, $ I $ and $\gamma_0$ on the proton. 
The test shows that it remains difficult for $\chi$EFT to  precisely describe {\it all}
observables in which spin degrees of freedom are explicit.   It provides strong
incentive for future improvements of calculations using $\chi$EFT, the leading
approach to the effective theory emerging directly from QCD, and for extending
the more fundamental lattice QCD calculations to the spin-dependent structure
of the nucleon.

\bigskip\noindent
{\bf{Methods}}

We measured   the spin difference yields 
  on the l.h.s. of  Eq.~(\ref{eq:polyield}) 
and solved that equation for $\Delta \sigma(W, Q^2)$, from which we extracted
$g_1$ and $A_1F_1$ as functions of $W$ and $Q^2$. 
We relied on the standard CLAS G{\scriptsize EANT}-3 Monte Carlo simulation package to fully simulate the spin-dependent yields, including all radiative effects and detector responses.
The efficiency of the modified 
Cherenkov Counter was determined by comparing data taken with only the Electromagnetic Calorimeter in the trigger
to those taken with the standard trigger that requires a coincidence between both detectors. 
The ratio of the latter to the former gave the Cherenkov efficiency. We selected only detector regions of well-understood acceptance in both the data and the simulation.
This process fully determined the function $a(W, Q^2)$ in Eq.~(\ref{eq:polyield}).  
  The same Eq.~(\ref{eq:polyield}) was also used to extract the
product $\mathcal{L}P_bP_t$ 
  by comparing
the measured yield difference (l.h.s. of Eq.~(\ref{eq:polyield})),
integrated over the elastic peak region 0.85 GeV $< W <$ 1.0 GeV, to the
   elastic cross section difference $\Delta \sigma(W = M, Q^2)$ 
  which can be calculated from 
the known electromagnetic form factors of the proton~\cite{Arrington:2007ux}.

The polarized cross section $\Delta \sigma(W, Q^2)$ in the simulation was calculated using
an event generator for inclusive electron scattering~\cite{Abe:1998wq} 
with up-to-date models of structure functions and asymmetries, including near-final  
data from JLab experiment E08-027.   
We extracted our results on $g_1$ and $A_1 F_1$ by varying our input 
parameterization for these quantities and finding the required values to make 
our simulation for the polarized yield agree with data. Corrections for higher-order quantum electromagnetic  effects (radiative corrections) were applied in the simulation, of which one effect is the high-energy tail from elastic scattering (elastic radiative tail).

We propagated the uncertainties on the polarized yields to the final values for $g_1$ and $A_1 F_1$. 
Systematic uncertainties were studied by changing model parameters, or other inputs, and re-running the simulation.  
  The overall uncertainty on the normalization factor 
 $\mathcal{L}P_bP_t$
for each beam energy varied
from 2\% to 5\%, dominated by the statistics of the measured
elastic peak and, to a lesser extent, the accuracy of the proton elastic form factors~\cite{Arrington:2007ux}
that enter into 
  $\Delta \sigma(W = M, Q^2)$ and hence into
our determination of   that factor.
Smaller contributions,
all less than 1\%, came from $\pi^-$ and $e^+e^-$ backgrounds, and scattering off the
slightly polarized $^{15}$N in the target. The reconstruction of $W$ has an uncertainty of less than 2~MeV, which was studied by shifting the simulated $W$ spectrum and repeating the extraction. 
Uncertainties due to trigger and particle reconstruction and identification inefficiencies,
as well as parameterizations for the structure functions, $F_{1,2}$ and $A_{1,2}$, were studied by varying them in the simulation. 
Uncertainties in the radiative corrections  
were estimated by varying the amount of material the electron passed through in the simulation, and by adjusting the elastic radiative tail within reasonable limits. 
  In all, the total experimental uncertainty is dominated by statistics.

To extrapolate our results on $ I(Q^2)$ to $Q^2=0$, we fit our data with a form obeying the $Q^2$-dependence of the  Alarc\'on {\it{et al.}} $\chi$EFT calculation~\cite{Alarcon:2020icz}. We chose this calculation 
because its $Q^2$-dependence agrees well with  our data over a wide $Q^2$ range.
We found the intercept of our fit with the $Q^2=0$ axis to be  $ I^{exp}(0)=-0.798\pm0.013\mathrm{(uncor)}\pm0.040\mathrm{(cor)} \pm{0.003}\mathrm{(range)} \pm0.003\mathrm{(form)}$, with  $\chi^2_\mathrm{reduced}=2.20$ determined with the ``uncor'' uncertainty.
Here, ``uncor'' and ``cor'' refer to the experiment point-to-point uncorrelated and
correlated uncertainties, respectively; ``range'' refers to the uncertainty due to the 
$Q^2$ range  ($Q^2\leq0.1$ GeV$^2$) used for the fit. 
The last contribution, ``form'', is the uncertainty on the $Q^2$-dependence used for the fit. It is calculated from the uncertainty band given by the $\chi$EFT calculation~\cite{Alarcon:2020icz}. 
Since the various uncertainties are largely independent, they are added quadratically, giving a total uncertainty of  $\pm 0.042$.  
  This is about twice smaller than that from photoproduction
measurements of $ I(0)$ because the $Q^2 \to 0$ extrapolation uncertainty calculated using~\cite{Alarcon:2020icz} is
negligible and because inclusive electroproduction automatically sums over all reaction channels, thereby removing uncertainties
associated with the detection of final states needed in photoproduction.
  On the other hand, the extrapolation uncertainty is calculated from~\cite{Alarcon:2020icz}, which disagrees with~\cite{Bernard:2012hb}. This
indicates that the uncertainty bands
provided in the calculations 
may not reflect the full theoretical uncertainties. 
Extrapolating using the $Q^2$-dependence from~\cite{Bernard:2012hb} yields 
$I^{exp}(0)=-0.625\pm 0.022$(uncor) $\pm 0.039$(cor) $\pm^{0.069}_{0.013}$(range) $\pm0.056$(form), with  $\chi^2_\mathrm{reduced}=2.23$ determined with the ``uncor'' uncertainty.
The ``uncor'' value here is larger
because the fit is limited to very few data points ($Q^2\leq0.024$ GeV$^2$). 
This result differs notably from our main result,
as expected 
from the very different slope of~\cite{Bernard:2012hb}. 
This discrepancy exemplifies the importance of testing and improving $\chi$EFT calculations,   
 since well-controlled predictions would make electroproduction data very competitive for verifying the GDH sum rule and other real photon observables.

\bigskip\noindent
{\bf{Data availability}}
Experimental data that support the findings of this study will be posted on the CLAS database, https://clasweb.jlab.org/physicsdb/, or are available from the corresponding author upon request.

\bigskip\noindent
{\bf{Code availability}}
The computer codes that support the plots within this paper and the findings
of this study are available from X. Zheng upon request.

\bigskip\noindent
{\bf{Author contributions}}
The members of the Jefferson Lab CLAS Collaboration constructed and operated the experimental
equipment used in this experiment. A large number of collaboration members participated
in the data collection. 
The following authors provided various contributions to the experiment design and commissioning, data processing, data analysis and Monte Carlo simulations: M. Battaglieri, R. De Vita, V. Drozdov, L. El Fassi, H. Kang, K. Kovacs, E. Long, M. Osipenko, S. Phillips, K. Slifer. 
The authors who performed the final data analysis and Monte Carlo
simulations were A. Deur, S.E. Kuhn, M. Ripani, J. Zhang, and X. Zheng. 

\noindent
The manuscript was reviewed by the entire CLAS collaboration before publication, and all authors approved
the final version of the manuscript.

\bigskip\noindent
{\bf{Competing interests}}
The authors declare no competing interests.

\bigskip\noindent
{\bf{Additional information}}

\noindent
Supplementary information are available online that includes all numerical results reported here. 

\noindent
Correspondence and requests for materials should be addressed to A.~Deur.

\noindent
Reprints and permissions information is available at ... (to be updated upon publication)

\acknowledgments{
We thank the personnel of Jefferson Lab for their efforts that 
resulted in the successful completion of the experiment. 
We are grateful to U.-G. Mei{\ss}ner and V. Pascalutsa for useful
discussions on the theoretical $\chi$EFT calculations. 
This work was supported by the U.S. Department of Energy 
(DOE), the U.S. National Science Foundation, the U.S. Jeffress Memorial Trust; 
the United Kingdom Science and Technology Facilities Council (STFC), the Italian Istituto Nazionale 
di Fisica Nucleare; the French Institut National de Physique Nucl\'eaire et de 
Physique des Particules, the French Centre National de la Recherche Scientifique; 
and the National Research Foundation of Korea. This material is based 
upon work supported by the U.S. Department of Energy, Office of Science, 
Office of Nuclear Physics under contract DE-AC05-06OR23177. }

\section*{References}

\newpage

{\bf \large Supplementary Material}

Results for $g_1^p$ and $A_1^pF_1^p$, along with their
  statistical and systematic uncertainties, are shown in the table and figure below.

\scriptsize{
\begin{longtable}{|c|c|c|c|c|c|c|c|} \hline

$Q^{2}$ & W  & $g_{1}$ & Stat. & Syst. & $A_{1}F_{1}$ & Stat. & Syst.\\
(GeV$^{2}$) & (GeV) &  & & &  & & \tabularnewline

 \hline
0.0120 & 1.11 & 0.1153 & 0.0898 & 0.0133 & 0.1448 & 0.1203 & 0.0172\tabularnewline
0.0120 & 1.13 & -0.0852 & 0.1148 & 0.0212 & -0.1189 & 0.1441 & 0.0262\tabularnewline
0.0120 & 1.15 & -0.1076 & 0.1197 & 0.0215 & -0.1458 & 0.1438 & 0.0272\tabularnewline
0.0120 & 1.17 & -0.2045 & 0.1168 & 0.0208 & -0.2479 & 0.1368 & 0.0270\tabularnewline
0.0120 & 1.19 & -0.7447 & 0.1360 & 0.0381 & -0.8537 & 0.1554 & 0.0472\tabularnewline
0.0120 & 1.21 & -0.6129 & 0.1437 & 0.0344 & -0.6873 & 0.1610 & 0.0436\tabularnewline
\end{longtable}


\begin{sidewaysfigure}[!ht]
\includegraphics[scale=0.37]{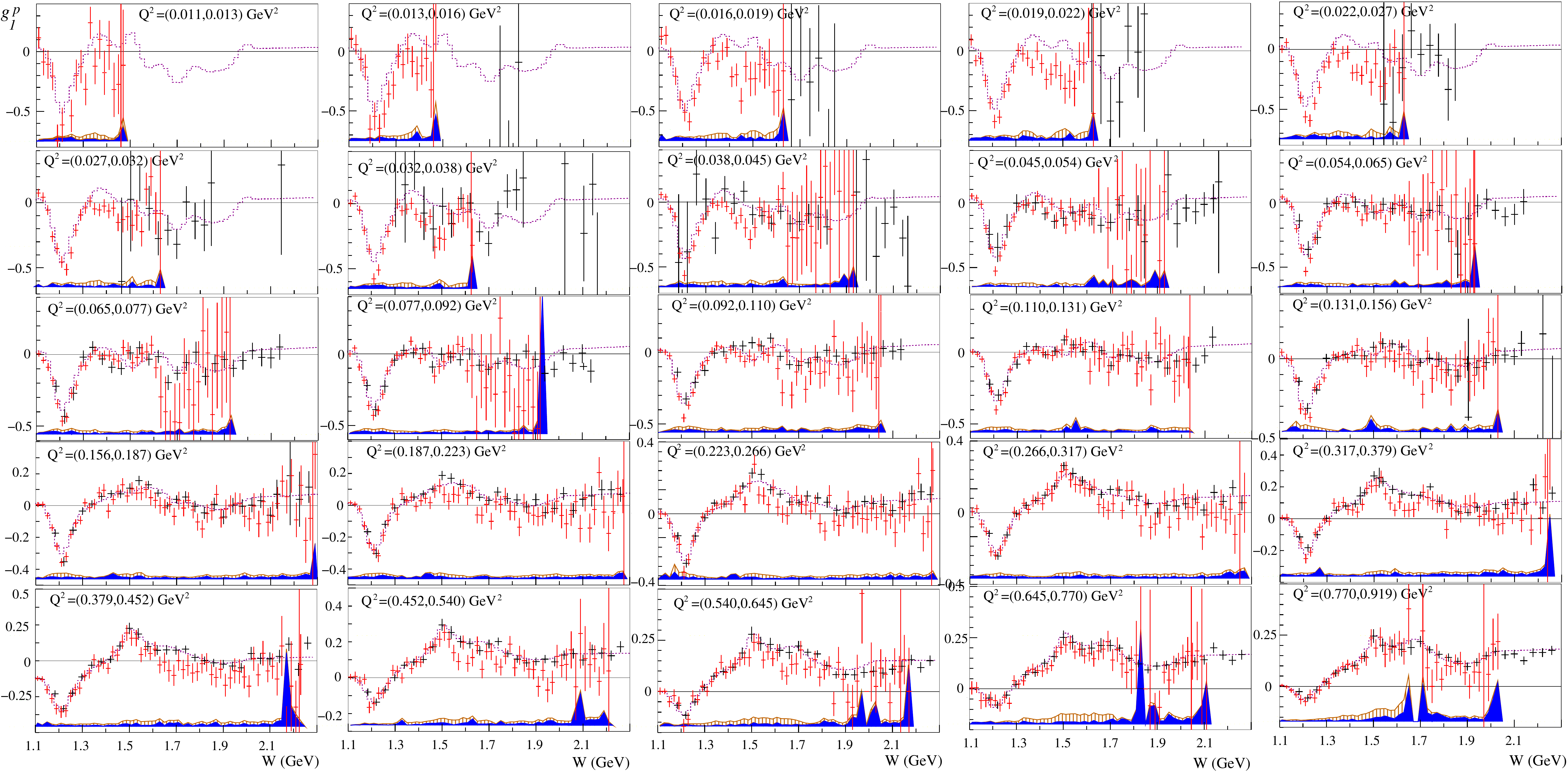}
\caption{Results on $g_1(Q^2,W)$ of the proton. Data from this work (red crosses) are plotted vs invariant mass $W$. 
  The error bars are statistical. The blue solid and the brown hatched bands show the 
  experimental and the parameterization uncertainties, respectively. The horizontal line is the zero
  of the vertical axis.
   The results are compared to
   a parameterization of previous world data (purple dotted curve) and results from a previous
   experiment carried out in Jefferson Lab's Hall B (black crosses, see Fersch, R. et al. (CLAS Collaboration). ``Determination of the proton spin structure functions for $0.05 < Q^2 < 5$ GeV$^2$ using CLAS.'' Phys. Rev. C {\bf96}, 065208 (2017). 
   The corresponding error bars display the statistical and systematic uncertainties added in quadrature).
}
\end{sidewaysfigure}

\newpage

Results for $\Gamma_1^p$, $I_{TT}^p$ and $\gamma_0^p$, for the measured and the full ranges, along with their
statistical and systematic uncertainties, are shown in the table below.
For Figs.~4 and 5 of the main article, the four lowest $Q^2$ bins were combined into two for visual clarity.
\begin{table}[!h]
\begin{center}
\scriptsize{
\begin{longtable}{|c|c|c|c|c|c|c|c|c|c|c|c|c|}
\hline 
$Q^{2}$ (GeV$^{2}$) & $\Gamma_1^{\mbox{data}}$ & $\Gamma_1^{\mbox{full}}$ & Stat. & Syst. & $I_{TT}^{\mbox{data}}$ & $I_{TT}^{\mbox{full}}$ & Stat. & Syst.& $\gamma_0^{\mbox{data}}$ & $\gamma_0^{\mbox{full}}$ & Stat. & Syst.\tabularnewline
\hline 
0.01200 & -0.00606 & -0.00615 & 0.00067 & 0.00060 & -1.00582 & -1.02906 & 0.10882 & 0.09780 & -2.47405 & -2.24818 & 0.29260 & 0.19167\tabularnewline
0.01430 & -0.00621 & -0.00631 & 0.00068 & 0.00061 & -0.87546 & -0.89909 & 0.09502 & 0.08632 & -1.96056 & -1.74642 & 0.26029 & 0.17095\tabularnewline
0.01708 & -0.00684 & -0.00695 & 0.00056 & 0.00020 & -0.82468 & -0.84534 & 0.06627 & 0.03565 & -1.72448 & -1.52370 & 0.17821 & 0.14287\tabularnewline
0.02042 & -0.00839 & -0.00848 & 0.00053 & 0.00021 & -0.86917 & -0.88748 & 0.05322 & 0.03746 & -1.87527 & -1.68530 & 0.13228 & 0.14971\tabularnewline
0.02436 & -0.01034 & -0.01041 & 0.00056 & 0.00026 & -0.92459 & -0.94052 & 0.04794 & 0.04096 & -1.97332 & -1.79514 & 0.11755 & 0.14808\tabularnewline
0.02904 & -0.00939 & -0.00942 & 0.00057 & 0.00023 & -0.74663 & -0.75999 & 0.04243 & 0.04423 & -1.66956 & -1.50446 & 0.10261 & 0.15216\tabularnewline
0.03466 & -0.01223 & -0.01222 & 0.00059 & 0.00027 & -0.83188 & -0.84289 & 0.03793 & 0.05111 & -1.71235 & -1.56216 & 0.09042 & 0.15319\tabularnewline
0.04139 & -0.01330 & -0.01219 & 0.00119 & 0.00029 & -0.78198 & -0.74516 & 0.05675 & 0.06003 & -1.52151 & -1.38013 & 0.08881 & 0.15918\tabularnewline
0.04940 & -0.01580 & -0.01452 & 0.00125 & 0.00029 & -0.80819 & -0.77233 & 0.05130 & 0.06886 & -1.53158 & -1.40976 & 0.08275 & 0.16844\tabularnewline
0.05902 & -0.01734 & -0.01585 & 0.00126 & 0.00033 & -0.79260 & -0.75820 & 0.04572 & 0.08078 & -1.51354 & -1.41314 & 0.07896 & 0.18236\tabularnewline
0.07047 & -0.02007 & -0.01832 & 0.00139 & 0.00038 & -0.80305 & -0.76970 & 0.04306 & 0.09128 & -1.48134 & -1.40324 & 0.07218 & 0.19327\tabularnewline
0.08412 & -0.02153 & -0.01949 & 0.00151 & 0.00048 & -0.75899 & -0.72679 & 0.04065 & 0.09988 & -1.32838 & -1.27278 & 0.06609 & 0.19773\tabularnewline
0.10054 & -0.02225 & -0.01978 & 0.00114 & 0.00082 & -0.70545 & -0.67357 & 0.02990 & 0.10381 & -1.18781 & -1.15468 & 0.05551 & 0.19272\tabularnewline
0.12004 & -0.02057 & -0.01760 & 0.00137 & 0.00097 & -0.61338 & -0.58177 & 0.03149 & 0.09889 & -1.03759 & -1.02438 & 0.05450 & 0.17324\tabularnewline
0.14295 & -0.02094 & -0.01733 & 0.00162 & 0.00117 & -0.55224 & -0.52006 & 0.03263 & 0.09423 & -0.84772 & -0.84995 & 0.05461 & 0.15771\tabularnewline
0.17080 & -0.01870 & -0.01493 & 0.00143 & 0.00145 & -0.48370 & -0.45770 & 0.02516 & 0.08985 & -0.75206 & -0.76529 & 0.03745 & 0.13620\tabularnewline
0.20421 & -0.01750 & -0.01274 & 0.00188 & 0.00151 & -0.40164 & -0.37302 & 0.02981 & 0.08304 & -0.54667 & -0.56500 & 0.04026 & 0.11746\tabularnewline
0.24355 & -0.00989 & -0.00342 & 0.00205 & 0.00149 & -0.28522 & -0.24978 & 0.03004 & 0.05469 & -0.40617 & -0.42440 & 0.04118 & 0.07387\tabularnewline
0.29038 & -0.00484 & 0.00394 & 0.00196 & 0.00227 & -0.21676 & -0.17225 & 0.02240 & 0.04132 & -0.30540 & -0.31800 & 0.02417 & 0.05353\tabularnewline
0.34662 & 0.00274 & 0.01414 & 0.00218 & 0.00228 & -0.12949 & -0.08033 & 0.02177 & 0.01918 & -0.18437 & -0.19751 & 0.02091 & 0.02428\tabularnewline
0.41389 & 0.00368 & 0.01938 & 0.00243 & 0.00299 & -0.12080 & -0.05993 & 0.02064 & 0.01856 & -0.15301 & -0.15964 & 0.01668 & 0.01485\tabularnewline
0.49404 & 0.01583 & 0.03663 & 0.00260 & 0.00373 & -0.02454 & 0.04532 & 0.01870 & 0.02476 & -0.05744 & -0.06060 & 0.01303 & 0.00869\tabularnewline
0.59017 & 0.02162 & 0.04879 & 0.00261 & 0.00524 & 0.00671 & 0.08231 & 0.01535 & 0.03136 & -0.02282 & -0.02647 & 0.00861 & 0.00962\tabularnewline
0.70473 & 0.02468 & 0.06261 & 0.00293 & 0.00684 & 0.01382 & 0.10601 & 0.01575 & 0.03110 & -0.01283 & -0.01191 & 0.00840 & 0.00785\tabularnewline
0.84121 & 0.02014 & 0.07757 & 0.00227 & 0.00370 & 0.01567 & 0.13133 & 0.00870 & 0.02078 & -0.00562 & -0.00166 & 0.00285 & 0.00447\tabularnewline
\hline
  \end{longtable}
}
\end{center}
\end{table}


\begin{thebibliography}{99}
%
\bibitem{Deur:2016tte} 
  Deur, A., Brodsky, S.J. \& de T\'eramond, G.F. 
  The QCD running coupling. 
       {{\it Prog.\ Part.\ Nucl.\ Phys.} {\bf 90}, 1--74 (2016).} 

\bibitem{Tanabashi:2018oca} 
Tanabashi, M. {et al.}, (Particle Data Group)  
Review of particle physics.  
     {{\it Phys.\ Rev.\ D} {\bf 98}, no. 3, 030001 (2018)}

\bibitem{Bernard:2007zu} 
  Bernard, V., 
  Chiral perturbation theory and baryon properties. 
       {{\it Prog.\ Part.\ Nucl.\ Phys.}  {\bf 60}, 82--160 (2008)}
  
\bibitem{Scherer:2009bt} 
  Scherer, S.,  
  Chiral perturbation theory: introduction and recent results in the one-nucleon sector. 
       {{\it Prog.\ Part.\ Nucl.\ Phys.}  {\bf 64}, 1--60 (2010)}

\bibitem{Kuhn:2008sy} 
  Kuhn, S.E., Chen, J.-P. \& Leader,~E., 
  Spin structure of the nucleon -- status and recent results. 
       {{\it Prog.\ Part.\ Nucl.\ Phys.} {\bf 63}, 1--50 (2009)} 

 \bibitem{Deur:2018roz} 
Deur, A., Brodsky, S.J. \& de T\'eramond, G.F. 
The spin structure of the nucleon. 
     {{\it Rep. Prog. Phys.} {\bf 82}, 076201 (2019)} 

\bibitem{Wu:1957my}
Wu, C., Ambler, E., Hayward, R., Hoppes, D. \& Hudson, R. 
Experimental test of parity conservation in beta decay. 
     {{\it Phys.\ Rev.} {\bf 105}, 1413--1414 (1957)}


\bibitem{Ellis:1973kp} 
  Ellis,~J.R. \& Jaffe,~R.L. 
A sum rule for deep inelastic electroproduction from polarized protons. 
     {{\it Phys.\ Rev.\ D} {\bf 9}, 1444--1446 (1974)}; [Erratum:
            {{\it Phys.\ Rev.\ D} {\bf 10}, 1669--1670 (1974)]}

\bibitem{Brodsky:1994kg} 
  Brodsky, S.J., Burkardt, M. \& Schmidt, I. 
QCD constraints on the shape of polarized quark and gluon distributions. 
     {{\it Nucl.\ Phys.\ B} {\bf 441}, 197--214 (1995)}


\bibitem{Chang:2018uxx} 
  Chang, C.C. {et al.},
  A percent-level determination of the nucleon axial coupling from quantum chromodynamics. 
       {{\it Nature} {\bf 558}, 91--94 (2018)}


\bibitem{Crabb:1995xi} 
  Crabb,~D.G. \& Day,~D.B. 
  The Virginia/Basel/SLAC polarized target: operation and performance during experiment E143 at SLAC. 
       {{\it Nucl.\ Inst.\ Meth.\ A} {\bf 356}, 9--19 (1995)}

\bibitem{Keith:2003ca} 
  Keith, C.D. {et al.} 
  A polarized target for the CLAS detector. 
       {{\it Nucl.\ Inst.\ Meth.\ A} {\bf 501}, 327--339 (2003)}


\bibitem{Mecking:2003zu} 
  Mecking,~B.A. {et al.} (CLAS Collaboration) 
  The CEBAF large acceptance spectrometer (CLAS). 
       {{\it Nucl.\ Inst.\ Meth.\ A} {\bf 503}, 513--553 (2003)}

\bibitem{Fersch:2017qrq} 
  Fersch, R. {et al.} (CLAS Collaboration). 
  Determination of the proton spin structure functions for $0.05<Q^2<5$ GeV$^2$ using CLAS. 
       {{\it Phys.\ Rev.\ C} {\bf 96},  065208 (2017)} 

\bibitem{Amarian:2002ar} 
  Amarian,~M. {et al.},
  The $Q^2$ evolution of the generalized Gerasimov-Drell-Hearn integral for the neutron using a $^3$He target. 
       {{\it Phys.\ Rev.\ Lett.}  {\bf 89}, 242301 (2002)}

\bibitem{Amarian:2003jy} 
 Amarian,~M. {et al.} (Jefferson Lab E94-010 Collaboration) 
 $Q^2$ evolution of the neutron spin structure moments using a $^3$He target. 
      {{\it Phys.\ Rev.\ Lett.}  {\bf 92}, 022301 (2004)} 

\bibitem{Amarian:2004yf} 
 Amarian,~M.{et al.} (Jefferson Lab E94-010 Collaboration)
 Measurement of the generalized forward spin polarizabilities of the neutron. 
      {{\it Phys.\ Rev.\ Lett.}  {\bf 93}, 152301 (2004)};

\bibitem{Deur:2004ti} 
  Deur,~A. {et al.}
  Experimental determination of the evolution of the Bjorken integral at low $Q^2$. 
       {{\it Phys.\ Rev.\ Lett.}  {\bf 93}, 212001 (2004)} 

\bibitem{Deur:2008ej} 
  Deur, A. {et al.}
  Experimental study of isovector spin sum rules. 
       {{\it Phys.\ Rev.\ D} {\bf 78}, 032001 (2008)} 

\bibitem{Dharmawardane:2006zd} 
  Dharmawardane, K.V. {et al.} (CLAS Collaboration)
  Measurement of the $x$- and $Q^2$-dependence of the asymmetry $A_1$ on the nucleon. 
       {{\it Phys.\ Lett.\ B} {\bf 641}, 11--17 (2006)};

\bibitem{Prok:2008ev} 
  Prok, Y. {et al.} (CLAS Collaboration)
  Moments of the spin structure functions $g_1^p$ and $g_1^d$ for
  $0.05 < Q^2 < 3.0$ GeV$^2$. 
       {{\it Phys.\ Lett.\ B} {\bf 672}, 12--16 (2009)}

  \bibitem{Guler:2015hsw} 
  Guler,~N. {et al.} (CLAS Collaboration)
  Precise determination of the deuteron spin structure at low to moderate $Q^2$ with CLAS and extraction of the neutron contribution. 
       {{\it Phys.\ Rev.\ C} {\bf 92}, 055201 (2015)} 

\bibitem{Chambers:2017dov} 
  Chambers,~A.J. {et al.}
  Nucleon structure functions from operator product Expansion on the lattice. 
       {{\it Phys.\ Rev.\ Lett.}  {\bf 118}, 24 242001 (2017)}

\bibitem{Liang:2019frk} 
  Liang,~J, Draper, T., Liu, K.-F., Rothkopf, A. \& Yang, Y.-B.
  (XQCD Collaboration) 
  Towards the nucleon hadronic tensor from lattice QCD.
       {{\it Phys.\ Rev.\ D} {\bf 101}, 114503 (2020)} 

\bibitem{Gerasimov:1965et} 
  Gerasimov, S.B.
  A sum rule for magnetic moments and the damping of the nucleon magnetic moment in nuclei. 
  {\it Sov.\ J.\ Nucl.\ Phys.}  {\bf 2}, 430--433 (1966)
  [{\it Yad.\ Fiz.} {\bf 2}, 598--602 (1965)]
  
\bibitem{Drell:1966jv} 
  Drell,~S.D. \& Hearn,~A.C. 
  Exact sum rule for nucleon magnetic moments. 
       {{\it Phys.\ Rev.\ Lett.} {\bf 16}, 908--911 (1966)}
    
  \bibitem{Dutz:2004zz} 
  Dutz,~H. {et al.} 
  Experimental check of the Gerasimov-Drell-Hearn sum rule for $^1$H. 
       {{\it Phys.\ Rev.\ Lett.}  {\bf 93}, 032003 (2004)}
    
 \bibitem{Hoblit:2008iy} 
  Hoblit,~S. {et al.} (LSC Collaboration) 
  Measurements of $\vec H \vec D (\vec \gamma, \pi)$ and implications for convergence of the Gerasimov-Drell-Hearn integral. 
       {{\it Phys.\ Rev.\ Lett.} {\bf 102}, 172002 (2009)}

\bibitem{Ji:1999mr} 
  Ji, X.-D. \& Osborne, J. 
  Generalized sum rules for spin-dependent structure functions of the nucleon. 
       {{\it J.\ Phys.\ G} {\bf 27}, 127--146 (2001)}

\bibitem{Bernard:1992nz} 
  Bernard, V., Kaiser, N. \& Meissner,~U.G. 
  Small momentum evolution of the extended Drell-Hearn-Gerasimov sum rule. 
       {{\it Phys.\ Rev.\ D} {\bf 48}, 3062--3069 (1993)}
   
\bibitem{Ji:1999pd} 
  Ji,~X.D., Kao, C.-W. \& Osborne, J.
  Generalized Drell-Hearn-Gerasimov sum rule at order O(p$^4$) in chiral perturbation theory.
       {{\it Phys.\ Lett.\ B} {\bf 472}, 1--4 (2000)}

\bibitem{Ji:1999sv} 
  Ji,~X.D., Kao, C.-W. \& Osborne, J.
  The Nucleon spin polarizability at order O(p$^4$) in chiral perturbation theory. 
       {{\it Phys.\ Rev.\ D} {\bf 61}, 074003 (2000)}

\bibitem{Bernard:2002bs} 
  Bernard, V., Hemmert,~T.R. \& Meissner, U.G. 
  Novel analysis of chiral loop effects in the generalized Gerasimov-Drell-Hearn sum rule. 
       {{\it Phys.\ Lett.\ B} {\bf 545}, 105--111 (2002)} 

\bibitem{Bernard:2002pw} 
  Bernard, V., Hemmert,~T.R. \& Meissner, U.G. 
  Spin structure of the nucleon at low energies. 
       {{\it Phys.\ Rev.\ D} {\bf 67}, 076008 (2003)}

\bibitem{Kao:2002cp} 
  Kao, C.W., Spitzenberg,~T. \& Vanderhaeghen, M. 
  Burkhardt-Cottingham sum rule and forward spin polarizabilities in heavy baryon chiral perturbation theory. 
       {{\it Phys.\ Rev.\ D} {\bf 67}, 016001 (2003)}

\bibitem{Bernard:2012hb} 
  Bernard, V., Epelbaum, E., Krebs, H. \& Meissner, U.G. 
  New insights into the spin structure of the nucleon. 
       {{\it Phys.\ Rev.\ D} {\bf 87}, 054032 (2013)}

\bibitem{Alarcon:2020icz}
Alarc\'on, J.M., Hagelstein, F., Lensky, V. \& Pascalutsa, V. 
Forward doubly-virtual Compton scattering off the nucleon in chiral perturbation theory: II. spin polarizabilities and moments of polarized structure functions. 
{{\it Phys. Rev. D} {\bf 102}, 114026 (2020)}

\bibitem{Lensky:2014dda} 
  Lensky, V., Alarcon,~J.M. \& Pascalutsa, V. 
  Moments of nucleon structure functions at next-to-leading order in baryon chiral perturbation theory. 
       {{\it Phys.\ Rev.\ C} {\bf 90},  055202 (2014)} 

\bibitem{Lensky:2016nui} 
  Lensky, V., Pascalutsa, V., \& Vanderhaeghen, M. 
  Generalized polarizabilities of the nucleon in baryon chiral perturbation theory. 
       {{\it Eur.\ Phys.\ J.\ C} {\bf 77},  119 (2017)}

\bibitem{Drechsel:2002ar} 
  Drechsel, D., Pasquini,~B. \& Vanderhaeghen,~M. 
  Dispersion relations in real \& virtual Compton scattering. 
       {{\it Phys.\ Rept.} {\bf 378}, 99--205 (2003)}

\bibitem{Burkert:1993ya} 
  Burkert, V.D. \& Ioffe, B.L.
  Polarized structure functions of proton and neutron and the Gerasimov-Drell-Hearn and Bjorken sum rules. 
  {\it J.\ Exp.\ Th.\ Phys.}  {\bf 78}, 619--622 (1994)

\bibitem{Pasechnik:2010fg} 
  Pasechnik, R.S., Soffer,~J. \& Teryaev,~O.V.
  Nucleon spin structure at low momentum transfers. 
       {{\it Phys.\ Rev.\ D} {\bf 82}, 076007 (2010)}

\bibitem{Guichon:1995pu} 
  Guichon,~P.A.M., Liu,~G.Q. \& Thomas,~A.W. 
  Virtual Compton scattering and generalized polarizabilities of the proton. 
       {{\it Nucl.\ Phys.\ A} {\bf 591}, 606--638 (1995)}

\bibitem{Gurevich:2017cpt} 
  Gurevich,~G.M. \& Lisin,~V.P. 
  Measurement of the proton spin polarizabilities at MAMI. 
       {{\it Phys.\ Part.\ Nucl.}  {\bf 48}, 111--116 (2017).}

\bibitem{Adhikari:2017wox} 
  Adhikari,~K.P. {et al.} (CLAS Collaboration) 
  Measurement of the $Q^2$ dependence of the deuteron spin structure function $g_1$ and its moments at low $Q^2$ with CLAS. 
       {{\it Phys.\ Rev.\ Lett.}  {\bf 120}, 062501 (2018)}

\bibitem{Sulkosky:2019zmn}
 Sulkosky.~V. {et al.} (Jefferson Lab E97-110 Collaboration)
 Measurement of the $^3$He spin-structure functions and of neutron ($^3$He) spin-dependent sum rules at $0.035\le Q^2 \le 0.24$ GeV$^2$. 
 {\it Phys. Lett. B} {\bf 805}, 135428 (2020)

\bibitem{Anderson:1972pca}
 Anderson,~P.W. 
 More is different. 
 {\it Science} {\bf 177}, 393--396 (1972)

\bibitem{Arrington:2007ux} 
  Arrington, J., Melnitchouk, W. \& Tjon,~J.A. 
  Global analysis of proton elastic form factor data with two-photon exchange corrections. 
  {\it Phys.\ Rev.\ C} {\bf 76}, 035205 (2007)
  
\bibitem{Abe:1998wq} 
  Abe,~K. {et al.} (E143 Collaboration) 
  Measurements of the proton and deuteron spin structure functions $g_1$ and $g_2$. 
       {{\it Phys.\ Rev.\ D} {\bf 58}, 112003 (1998)}

\end{thebibliography}
\end{document}